\documentclass[aps,prl,reprint]{revtex4-1}

\usepackage{blindtext}
\usepackage{graphicx}
\usepackage{braket}
\usepackage{amsmath} 
\usepackage{amssymb}
\usepackage{bm}
\usepackage[separate-uncertainty=true, multi-part-units = single, product-units=single]{siunitx}
\usepackage{upgreek}
\usepackage{color}

\begin{document}
\title{A Rydberg-dressed Magneto-Optical Trap}
\author{A. D. Bounds}
\author{N. C. Jackson}
\author{R. K. Hanley}
\author{R. Faoro}
\author{E. M. Bridge}
\author{P. Huillery}
\author{M. P. A. Jones}
\email{m.p.a.jones@durham.ac.uk}
\affiliation{Joint Quantum Centre Durham-Newcastle, Department of Physics, Durham University, Durham UK }

\begin{abstract}
We propose and demonstrate the laser cooling and trapping of Rydberg-dressed Sr atoms. By off-resonantly coupling the excited state of a narrow ($\SI{7}{kHz}$) cooling transition to a high-lying Rydberg state, we transfer Rydberg properties such as enhanced electric polarizability to a stable magneto-optical trap operating at $<\SI{1}{\upmu K}$. By increasing the density to $\SI{1e12}{cm^{-3}}$, we show that it is possible to reach a regime where the long-range interaction between Rydberg-dressed atoms becomes comparable to the kinetic energy, opening a route to combining laser cooling with tunable long-range interactions.
\end{abstract}

\maketitle

The strong dipolar interactions between Rydberg atoms have been exploited to perform numerous experimental breakthroughs in many-body quantum simulation \cite{Lukin2001,Browaeys2016,Zeiher2016,Bernien2017}, quantum information processing \cite{Isenhower2010,Saffman2010,Wilk2010} and quantum non-linear optics \cite{Firstenberg2016,Adams2010,Dudin2012,Peyronel2012,Maxwell2013}. To take advantage of coherent quantum dynamics, most of these realizations have focused on timescales shorter than the lifetime of the Rydberg state. However, there is a growing interest in extending the investigation time of Rydberg ensembles for applications in quantum simulation and metrology \cite{Weimer2010,Keating2013,Gil2014}. A promising method is to off-resonantly couple the ground state to a Rydberg state, resulting in the controlled admixture of some interacting Rydberg character \cite{Santos2000,Bouchoule2002,Johnson2010,Honer2010,Glaetzle2012}. This so-called Rydberg dressing approach could enable the realization of supersolids \cite{Henkel2010,Pupillo2010,Cinti2010,Boninsegni2012}, frustrated or topological quantum magnetism \cite{Lee2013,Glaetzle2014,Glaetzle2015,Pohl2015} or spin squeezing for enhanced metrology \cite{Bouchoule2002,Gil2014}. Experimentally, Rydberg-dressing has been demonstrated for two atoms \cite{Jau2015} and in optical lattices \cite{Zeiher2016,Zeiher2017}, but it seems to be more challenging in randomly distributed ensembles due to uncontrolled loss mechanisms \cite{Aman2016,Goldschmidt2016,Helmrich2016,Boulier2017}.

\begin{figure}[h]
\centering
\includegraphics[width=\columnwidth]{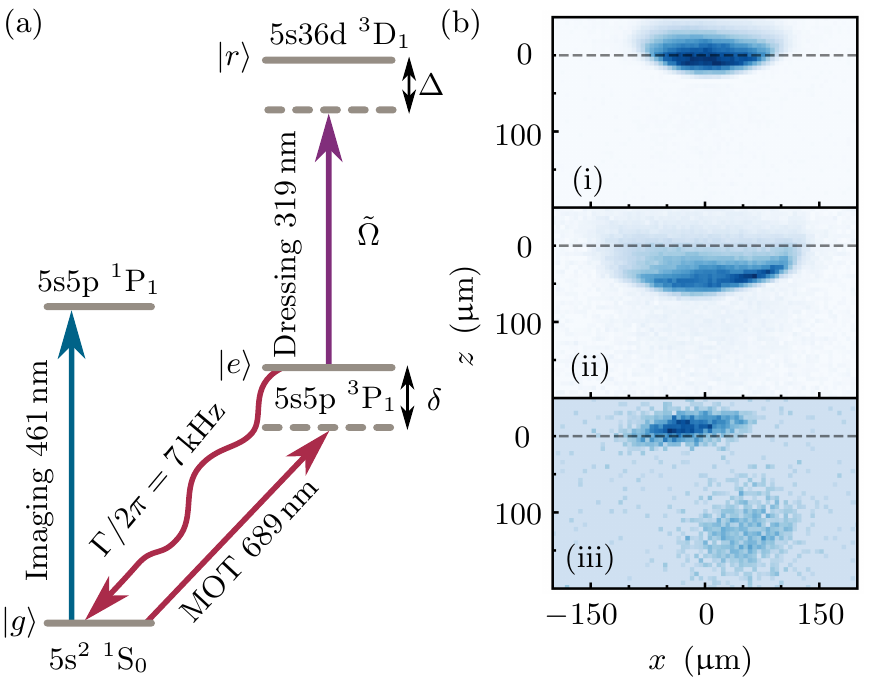}
\caption{\label{fig:fig1}
(a) A MOT operating on the $\SI{689}{nm}$ intercombination transition is dressed by off-resonantly coupling the state $\vert e \rangle$ to a Rydberg state $\vert r \rangle$. The strong $\SI{461}{nm}$ transition is used for absorption imaging.
(b) Images of (i) the undressed MOT with $\delta/2\pi = \SI{-110}{kHz}$, (ii) after 5 ms of dressing with $\delta/2\pi = \SI{-110}{kHz}$ (iii) after \SI{5}{ms} of dressing with the detuning altered to $\delta/2\pi = \SI[retain-explicit-plus]{+190}{kHz}$. Here $\tilde{\Omega}/2\pi = \SI{4}{MHz}$ and $\Delta/2\pi= \SI{12}{MHz}$. The dashed line indicates the position of the undressed MOT.} 
\end{figure} 

In this Letter we introduce a new scheme where Rydberg dressing is applied to an excited state undergoing spontaneous emission (Fig.~\ref{fig:fig1}). We show that Rydberg-dressed atoms can be laser cooled to sub-$\rm{\upmu}$K temperatures and trapped in a magneto-optic trap (MOT), while simultaneously acquiring the properties of the Rydberg state such as enhanced sensitivity to DC electric fields. The result is a hybrid magneto-electro-optical trap controllable by electric as well as magnetic fields. In addition we show that the Rydberg-dressed MOT can operate in a regime where the strength of the interactions is comparable to the dissipation and the kinetic energy, and with a lifetime that exceeds that of the Rydberg state by a factor of ${\sim}70$. Although laser cooling of Rydberg-dressed atoms has been proposed as a way to protect crystalline phase against dissipative effects \cite{Glaetzle2012},
active cooling of Rydberg gases is a relatively unexplored area \cite{Guo1996,Korsunsky1995} where interesting physics could arise from the presence of cooling and the mechanical effect of the interactions. Examples could include collisional Sisyphus-like cooling \cite{Vogl2009,Zhao2012} induced by the Rydberg-dressed potential or, in the spirit of anti-blockade experiments \cite{Amthor2010}, cooperative cooling where the collective scattering of multiple photons by groups of atoms dominates over single particle cooling. 

The laser cooling and Rydberg dressing scheme is shown in  Fig.~\ref{fig:fig1}(a). The goal is to dress the upper state $\vert e \rangle$ of the cooling transition by coupling it to a Rydberg state $\vert r \rangle$ with Rabi frequency $\tilde{\Omega}$ and detuning $\Delta$. In the weak dressing limit $\tilde{\Omega} \ll \Delta$, this creates a new Rydberg-dressed excited state $\vert \tilde{e} \rangle \approx \vert e \rangle - \epsilon \vert r \rangle$ with dressing fraction $\epsilon = \tilde{\Omega}/(2\Delta)$, that experiences an  AC Stark shift $\delta_\mathrm{AC} = \Delta(\sqrt{1+4\epsilon^2}-1)/2$. In contrast to previous experiments 
 \cite{Aman2016}, the detuning $\delta$ of the cooling laser is small, such that significant population is transferred to state $|\tilde{e}\rangle$. 

In addition to the AC Stark shift, two atoms in the dressed state $\vert \tilde{e} \rangle$ separated by a distance $r$ experience a soft-core dressed interaction potential \cite{Henkel2010,Pupillo2010} $V(r)=V_0(1+(r/R_c)^6)^{-1}$ with a peak magnitude $V_0 = \hbar \tilde{\Omega}^4/8 \vert \Delta \vert^3$ and length scale $R_c = \vert C_6/2\hbar\Delta \vert^{\frac{1}{6}}$. Here $C_6$ is the van der Waals coefficient associated with the Rydberg state $|r\rangle$. For reasonable choices of $\tilde{\Omega}$ and $\Delta$, $\delta_\mathrm{AC}$ and $V_0$ typically range from $0 \rightarrow \SI{100}{kHz}$, far smaller than the linewidth of most cooling transitions. 

We circumvent this limitation by using the $5\mathrm{s}^2\,^{1}\rm{S}_{0}\rightarrow 5\mathrm{s}5\mathrm{p}^{3}\,\rm{P}_{1}$ intercombination line in ${^{88}}$Sr with a linewidth of $\Gamma/2\pi = \SI{7}{kHz}$. Thus $V_0$  is comparable to both the linewidth and the kinetic energy $V_0 \approx \{\hbar\Gamma, k_{\rm{B}} T$\}, while the interaction length scale $R_{c}$ can also exceed the mean spacing between atoms in the excited state $\left\langle r_e \right\rangle$. Importantly, the behaviour of the non-interacting MOT, including the effects of additional AC Stark shifts, can be simulated to very high accuracy  \cite{Hanley2017}, enabling a quantitative understanding of the effects of Rydberg dressing.

The experiments begin with the formation of a MOT on the intercombination line; the loading procedure and apparatus  are  described in \cite{BoddyThesis,MillenThesis}. The MOT was operated with an intensity per beam of $I$ = 4-10$I_{\rm{sat}}$, where $I_{\rm{sat}}$ is the saturation intensity, resulting in  $1/e^{2}$ cloud radii of \SI{30}{\upmu m} (\SI{100}{\upmu m}) in the vertical ($z$) (horizontal ($x$)) direction, and temperature $T_\mathrm{z}=$\SI{800}{nK}.  The trap forms where the detuning $\delta$ of the cooling laser matches the Zeeman shift from the quadrupole magnetic field and the gravitational sag \cite{Loftus2004}. As a result the detuning seen by the atoms is largely fixed and instead $\delta$ determines the cloud shape and position with a sensitivity of ${\sim}$\SI{10}{kHz}. To form a Rydberg-dressed MOT, the excited state $\rm{\vert e \rangle = 5s5p \ ^{3}P_{1}}$ was coupled to the Rydberg state $\vert r \rangle = \rm{5s36d \ ^{3}D_{1}}$, for which the interactions are weakly attractive \cite{Vaillant2012}. The horizontally propagating dressing laser produced up to $\SI{250}{mW}$ at $\SI{319}{nm}$ \cite{Bridge2016}, and was linearly polarized in the vertical ($z$) direction. The $1/e^{2}$ beam radius was $\SI{120}{\upmu m}$ ($\SI{160}{\upmu m}$) in the horizontal (vertical) direction, and the associated Rabi frequency $\tilde{\Omega}$ was measured using Autler-Townes splitting. After dressing for a time $t_\mathrm{d}$, the cloud was imaged  at an angle of $30^{\circ}$ to the coupling beam via absorption on the 
$\SI{461}{nm}$ transition.

The effect of the dressing laser is shown in Fig.~\ref{fig:fig1}(b). For the chosen parameters $\epsilon^2 = 0.03$  and $\bar{V}_0 \equiv V_0/(\hbar\Gamma) = 2.5 $. The peak number of atoms per dressed blockade sphere $N_\mathrm{c} = \eta \rho 4 \pi R_\mathrm{c}^3/3 = 0.5$, where $\eta=0.05$ is the fraction of atoms in the state $|\tilde{e}\rangle$ (see Supplementary Information), such that the MOT is in the non-interacting regime. Fig.~\ref{fig:fig1}(b(i-ii)) show that the primary consequence of the dressing is a significant vertical shift of the MOT, since the AC Stark shift $\delta_{\mathrm{AC}}$ adds to the detuning $\delta$ experienced by the atoms. By simultaneously adjusting the detuning such that $\delta\rightarrow \delta -\delta_\mathrm{AC}$ during the dressing stage we compensate this AC Stark shift, and the MOT remains at its original position (Fig.~\ref{fig:fig1}(b(iii))). The vertical shift is eliminated for a detuning compensation of \SI[retain-explicit-plus]{+300}{kHz}, close to the calculated peak AC Stark shift of \SI[retain-explicit-plus]{+325}{kHz}. Importantly, the cooling laser is now blue-detuned with respect to the bare transition. Therefore the compensated MOT only traps Rydberg-dressed atoms; undressed atoms outside the dressing beam are observed falling away  (Fig.~\ref{fig:fig1}(b(iii))). 

\begin{figure}
\centering
\includegraphics[width=\columnwidth]{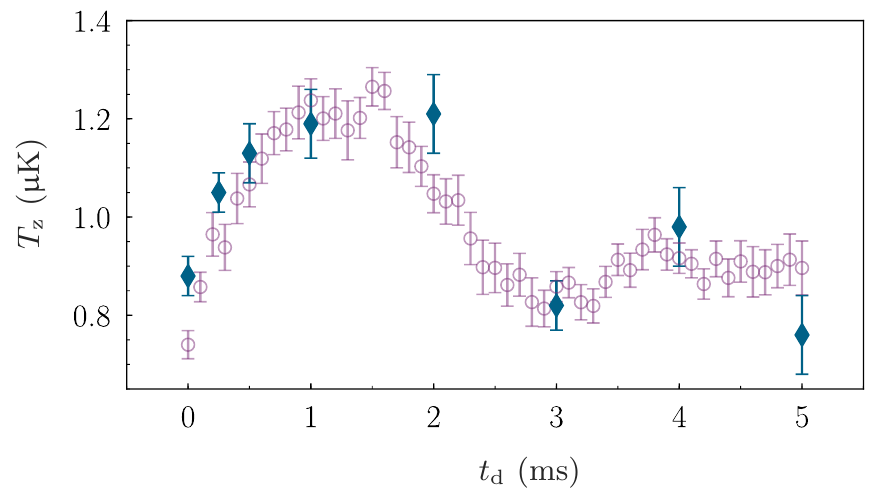}
\caption{\label{fig:fig2}
Measured (blue diamonds) and simulated (purple circles) temperature in the vertical ($z$) direction $T_\mathrm{z}$ versus dressing time $t_{\rm{d}}$ for $\tilde{\Omega}/2\pi = \SI{4}{MHz}$, $\Delta/2\pi= \SI{12}{MHz}$ and $\delta/2\pi = \SI{190}{kHz}$. Errorbars indicate the standard error of the mean.}  
\end{figure}

To verify that the Rydberg atoms are laser cooled as well as trapped, we measured the temperature (using ballistic expansion) as a function of $t_{\rm{d}}$ as shown in Fig.~\ref{fig:fig2}. Heating is observed in the first millisecond, but the temperature subsequently returns to that of the initial undressed MOT, showing that laser cooling is active during the dressing. To gain further insight into this behaviour we adapted the quantitative Monte Carlo simulations described in \cite{Hanley2017}, adding the spatially-dependent AC Stark shift due to the dressing beam (see Supplementary Information). The simulation is in quantitative agreement with the data, where the only fit parameter is the position of the dressing beam relative to the center of the quadrupole magnetic field, in this case $\SI{-20}{\upmu m}$ below and $\SI{60}{\upmu m}$ to the side of the quadrupole centre. These simulations showed that the initial heating is the result of the spatial dependence of the AC Stark shift, which transiently leads to increased scattering from the MOT beams as atoms find themselves at the ``wrong'' detuning. We confirmed experimentally that this effect depends on the UV beam alignment, reflecting the sensitivity of the MOT to small changes in detuning. Subsequent cooling then occurs as the cloud shape adapts to match the new spatially-dependent resonance condition.
 
\begin{figure}[h]
\centering
\includegraphics[width=\columnwidth]{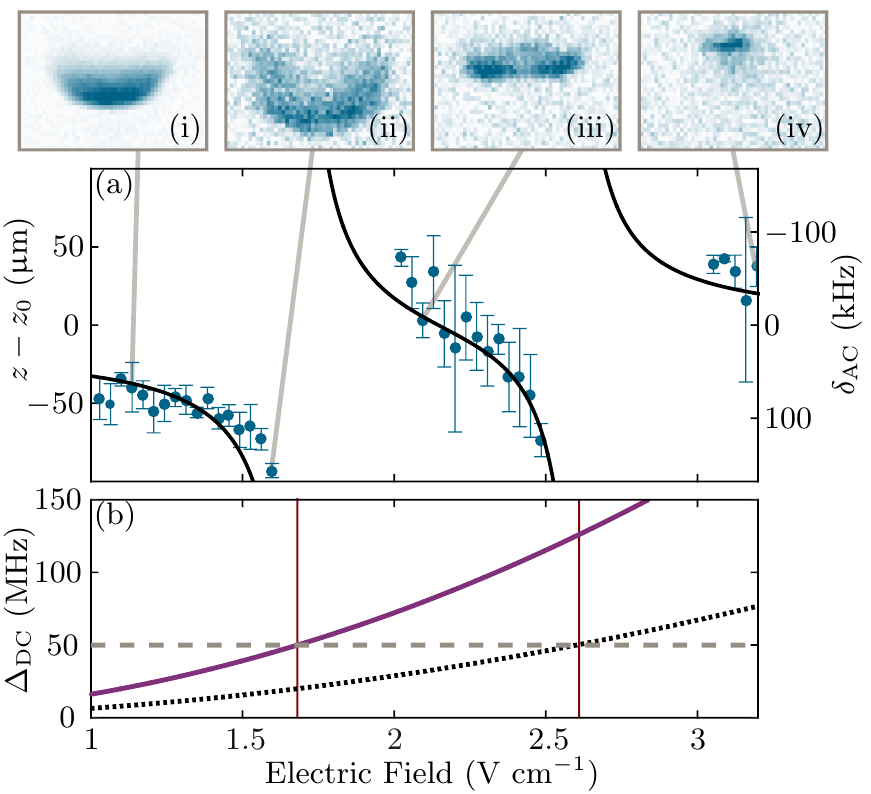}
\caption{\label{fig:fig3} (a) The measured (blue circles) and predicted (black line) change in vertical MOT versus electric field with $\rm{t_d}$=\SI{10}{ms}, $\tilde{\Omega}/2\pi$ = \SI{4}{MHz}, and $\delta/2\pi$ = \SI{-110}{kHz}. Images (i)-(iv) show the MOT at fields of 1.1, 1.6, 2.1 and 3.2 $\rm{V~cm^{-1}}$ respective;y. (b) DC Stark shift of the $|m_\mathrm{J}|=1$ (dotted black line) and $|m_\mathrm{J}|=0$ (purple line) components of the $\rm{36d \ ^{3}D_{1}}$ state. Dashed grey line indicates $\Delta/2\pi=50$\ MHz.
}
\end{figure}

An essential aspect of Rydberg dressing is that the dressed atoms acquire characteristics of the Rydberg state. To show that this can occur alongside cooling and trapping, we demonstrate that the dressed MOT becomes sensitive to an applied DC electric field as a consequence of the high polarizability $\alpha_\mathrm{r}$ of the Rydberg state. A static electric field  was applied in the horizontal ($x,y$) plane at an angle of $30^{\circ}$ to the coupling beam during the dressing time $t_\mathrm{d}=10$ ms using segmented ring electrodes \cite{MillenThesis}. As shown in Fig.~\ref{fig:fig3}(a), the applied electric field has a drastic effect on the shape, density and position of the MOT. At low field strength (Fig.~\ref{fig:fig3}(a(i))) where the DC Stark shift $\Delta_\mathrm{DC}$ is small, $\Delta> \{\tilde{\Omega}, \Delta_\mathrm{DC}\} $ and the dressed state picture remains valid. In this case, atoms in $|\tilde{e}\rangle$ acquire a polarizability given by $\alpha_\mathrm{e}= \epsilon^2 \alpha_\mathrm{r}$. For the parameters of Fig.~\ref{fig:fig3} $\alpha_\mathrm{e} \approx \SI{e-33}{Cm^2 V^{-1}}$ which is ${\sim}10^5$ times larger than that of the bare $\mathrm{5s5p}\,^3P_J$ states \cite{Middelmann2012,Sherman2012}. The resulting change in $\delta$ leads to an associated position shift. For larger fields $\tilde{\Omega} \approx \Delta-\Delta_{\mathrm{DC}}$ and the variation in AC Stark shift is associated with population transfer to $|r\rangle$, leading to loss (Fig.~\ref{fig:fig3}(a(ii))). The MOT distorts, as atoms in the center and wings of the cloud experience different coupling strengths due to the spatial inhomogeneity of the quadrupole field; this effect is particularly apparent for values of electric field where the dressing beam frequency lies between the resonances associated with the  $|m_\mathrm{J}|=0,1$ components (Fig.~\ref{fig:fig3}(a(iii))). At higher field strengths, the sign of the AC Stark shift is reversed (Fig.~\ref{fig:fig3}(a(iv))), and the MOT forms above its initial position. Away from resonances, the shift in the vertical position of the cloud is in good agreement with that predicted from the AC Stark shift, as shown in Fig.~\ref{fig:fig3}(a), provided the expected stronger coupling to the $|m_{\mathrm{J}}| = 0$ state due to the polarization of the MOT and dressing beams is taken into account.  These results highlight that a trap exists for all values of electric field for which the dressing beam is sufficiently far from resonance. The Rydberg-dressed MOT can thus be viewed as a novel type of hybrid trap whose size, shape and position can  be controlled by a relatively weak electric as well as magnetic fields. This enhanced sensitivity to spatially-dependent electric fields could find applications in electrometry for optical lattice clocks \cite{Bowden2017}.

\begin{figure}
\centering
\includegraphics[width=\columnwidth]{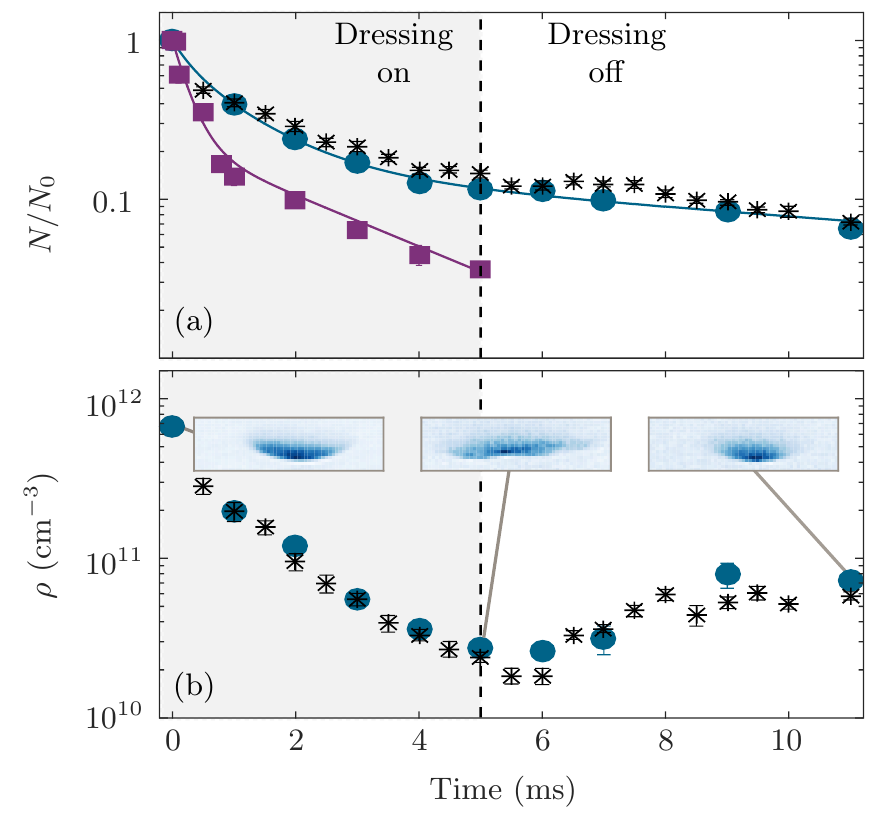}
\caption{\label{fig:fig4}
(a) Normalised atom number $N/N_\mathrm{t=0}$ as a function of time for the $\rm{5s36d ^{3}~D_{1}}$ (purple squares), $\rm{5s36d ^{3}~D_{2}}$  (blue circles) and $\rm{5s37d ^{3}~D_{2}}$ (black stars) Rydberg states. Solid lines represent double exponential fits. Shading indicates the region where the dressing laser is on (b) Variation of $\rho$ with time for the $\rm{5s36d ^{3}~D_{2}}$  (blue circles) and $\rm{5s37d ^{3}~D_{2}}$ (black stars) states. Here $\tilde{\Omega}/2\pi$ = \SI{4}{MHz}, $\Delta/2\pi$ = \SI{12}{MHz} ($\rm{5s36d\,^{3}D_{1}}$) and  $\tilde{\Omega}/2\pi$ = \SI{5}{MHz}, $\Delta/2\pi$ = \SI{30}{MHz} ($\rm{5s(36,37)d\, ^{3}D_{2}}$). Insets: images at the times indicated for the $\rm{5s36d ^{3}~D_{2}}$ state.}
\end{figure}

The observation of interactions requires sufficient interaction strength $\bar{V}_0 \approx 1$ and density ($N_\mathrm{c} > 1$) to be maintained over the $\sim \SI{3}{ms}$ timescale associated with the atomic motion (Fig.~\ref{fig:fig2}). Measurements of the atom number as function of $t_\mathrm{d}$ are shown in Fig.~\ref{fig:fig4}(a). For the $\rm{5s36d \ ^{3}D_{1}}$ state considered so far we observe rapid loss with a decay constant of $\SI{0.4 \pm 0.1}{ms}$ at short times. Using a micro-channel plate detector, we determined that this loss is associated with ionization \cite{BoundsThesis}. The DC Stark shift due to ions in the cloud brings the dressing laser onto resonance, leading to enhanced Rydberg excitation and loss. Although we observed that this effect can be partially suppressed by applying a DC electric field, a better approach is to switch to a Rydberg state with the opposite sign of DC Stark shift. The $\rm{5snd \ ^{3}D_{2}}$ states (for $n<37$) have repulsive van der Waals interactions and a positive Stark shift for $n<37$, such that ions shift nearby atoms out of resonance with the red-detuned dressing laser. For the $n=36,37$ states we clearly observe a reduced trap loss at short times with an electric field-independent lifetime of \SI{1.2\pm0.5}{ms} for $n=36$, which is ${\sim}70$ times larger than the measured lifetime of the Rydberg state ($\tau=$\SI{18\pm1}{\upmu s}). A drawback of working with the $\rm{5snd \ ^{3}D_{2}}$ states is that the change in sign of $\delta_\mathrm{AC}$ combined with the Gaussian spatial profile of the dressing beam leads to weaker confinement (see Supplementary Information), limiting the maximum interaction strength we could achieve to $\bar{V}_0=0.4$ ($N_\mathrm{c}=1.7$). This effect also appears to be responsible for the residual density-dependent decay at short times.  At $t_\mathrm{d} = 3$~ms 20\% of the atoms remain, and the subsequent decay (lifetime  \SI{5\pm1}{ms}) is compatible with the observed decay of the undressed MOT after the UV beam is switched off  (Fig.~\ref{fig:fig4}(a)). 

Surprisingly, the density (Fig.~\ref{fig:fig4}(b)) decays faster than the atom number. Images of the MOT (insets Fig.~\ref{fig:fig4}(b)) show that  this is due to a significant expansion of the cloud in the horizontal direction during dressing. After the dressing laser is switched off, the cloud returns to its original size and the density is largely restored. This expansion is independent of the principal quantum number over the range $n=34-37$ where we expect $N_\mathrm{c}$ to vary from $0.7\rightarrow 25$ \cite{ForsterNote}, indicating it is not caused by the dressed interaction. Instead it is due to the combined effect of radiation pressure forces \cite{Sesko1991} and the spatial inhomogeneity of the AC Stark shift (see Supplemental Material). 

The results shown in Fig.~\ref{fig:fig4} are encouraging. The achievable values for the lifetime, $\bar{V}_0$ and $N_\mathrm{c}$ are limited by the non-uniform intensity of the dressing beam, rather than by Rydberg excitation. A straightforward way to overcome these effects is to use additional beam shaping optics to produce a uniform intensity profile. With this improvement, the data in Fig.~\ref{fig:fig4} suggest that the necessary conditions for interactions can be maintained for ${\sim}3$ ms. So far we have not observed excess loss due to Rydberg impurities \cite{Aman2016,Goldschmidt2016,Boulier2017}, perhaps because the spontaneous decay of $|\tilde{e}\rangle$ naturally realises the stroboscopic dressing scheme proposed to overcome the loss in ground-state dressing experiments \cite{Aman2016,Boulier2017,Gaul2016}. 

Lastly we turn our attention to possible signatures of Rydberg-dressed interactions. Previous work has considered the motion in the fully-coherent regime \cite{Macri2014,Genkin2014,Buchmann2017}, or alternatively the effects of dissipation in the frozen gas regime where motion is neglected \cite{Weimer2010,Malossi2014,Marcuzzi2014}. However the Rydberg-dressed MOT operates in a complex regime where the interaction strength, dissipation and kinetic energy are all comparable in scale, limiting the usefulness of common approximations for such many-body systems. A possible solution is to combine classical equations of motion with a quantum jump approach for the internal states \cite{Glaetzle2012,Sibalic2016}, although for the many-body case $N_\mathrm{c}>1$ this appears demanding. 

\begin{figure}
\centering
\includegraphics[width=\columnwidth]{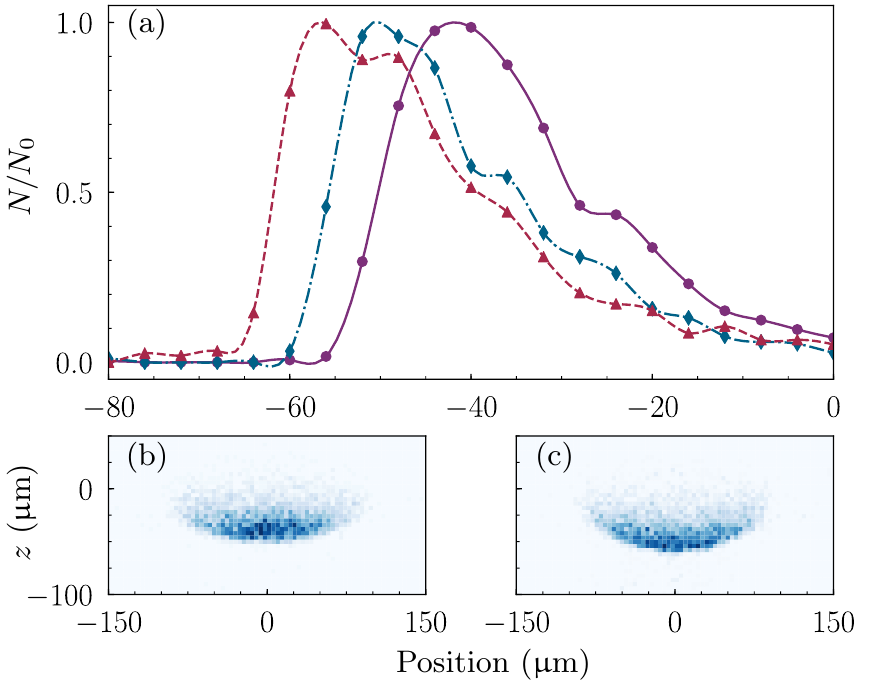}
\caption{\label{fig:fig5} (a) Vertical slices through the center ($x=0$) of the simulated density distribution for parameters: non-interacting ($\bar{V}_0=0$ (purple circles)), $\{\Omega/2\pi = \SI{16}{MHz}, \Delta/2\pi = \SI{-96}{MHz}, \bar{V}_0=1.25\}$ (blue diamonds), $\{\Omega/2\pi = \SI{16}{MHz}, \Delta/2\pi = \SI{-68}{MHz}, \bar{V}_0 =3.5 \}$ (red triangles). The lines are a guide to the eye. Images show the full 2D density distribution for the (b) non-interacting ($\bar{V}_0=0$) (c) interacting ($\bar{V}_0 =3.5$) clouds.}
\end{figure}

Instead we considered a more straightforward mean-field approach. The two-dimensional model (see Supplemental Material) accounts for many-body nature of the dressed interaction by using a step-function approximation to the dressed potential \cite{Bouchoule2002,Balewski2014}.
The result is an additional detuning that depends on the local density of the cloud. By adding this to our  model, we recursively solve for the density distribution in the presence of interactions and one-body loss from the Rydberg state. The results are shown in Fig.~\ref{fig:fig5} for a uniform  intensity $300\times100$~{$\upmu$m}   dressing beam  with power \SI{1}{W}, for Rydberg fractions $ \epsilon^2 = 0.007$ as in Fig.~\ref{fig:fig4} and for $\epsilon^2 =0.014$. The interaction causes a significant density-dependent shift in the vertical position of the cloud, associated with a ``bending'' of the cloud since the effect of interactions is larger in the center than in the wings due to the density-dependence of the mean-field interaction. Both effects are large enough to be observed in future experiments. We note that beyond mean-field effects due to correlations are also expected, since the Rydberg-dressed interaction strongly correlates the scattering rate and hence the cooling of neighboring atoms.

In summary we have demonstrated the operation of a Rydberg-dressed magneto-optical trap operating with a lifetime in excess of \SI{1}{ms} and active laser cooling to sub-$\upmu$K temperatures. The admixed Rydberg fraction is evidenced by the creation of a hybrid magneto-electro-optical-trap that exploits both the large static polarizability of the Rydberg state and the long lifetime afforded by Rydberg dressing. The achievable interaction strength and density is currently limited by the non-uniform spatial profile of the dressing beam, which is straightforward to overcome in future experiments. A simple model predicts that with these changes, the effect of the Rydberg-dressed interaction should lead to observable changes to the cloud shape. As such, this work represents a promising step towards combining laser cooling with tunable long-range interactions.

Financial support was provided by EPSRC grant EP/J007021/. This project has also received funding from the European Union's Seventh Framework (grant agreement no. 612862-HAIRS) and Horizon 2020 (grant agreements nos. 660028-EXTRYG and  640378-RYSQ) research and innovation programs. The data presented in this paper are available for download (to add). A. D. Bounds and N. C. Jackson contributed equally to this work.

\begin{center}
\textbf{Supplemental Material}
\end{center}

\subsection{Modelling the dressed MOT}
\begin{figure}
\centering
\includegraphics[width=\columnwidth]{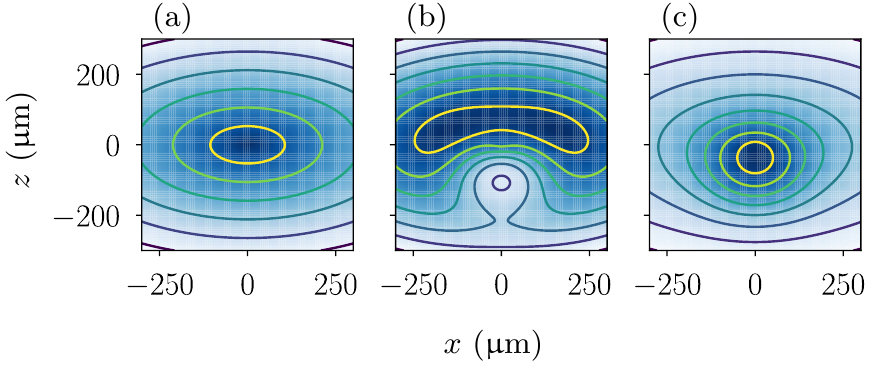}
\caption{\label{fig:Res_cond}
The figure shows contours of constant energy of the $m_j = -1$ state in the presence of the magnetic quadrupole field, with a vertical gradient of $\SI{8}{G cm^{-1}}$, for the case of no dressing (a) and dressing with $\Delta<0$ (b) or $\Delta>0$ (c) respectively.}
\end{figure}

\begin{figure*}
\centering
\includegraphics[width=\textwidth]{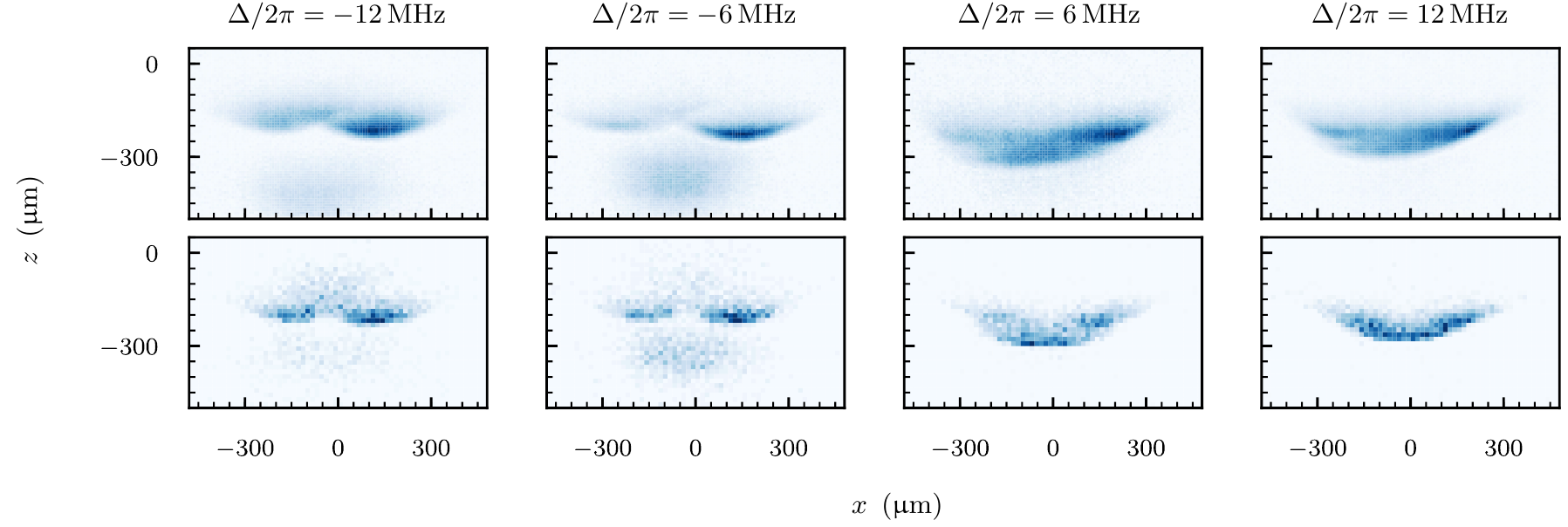}
\caption{\label{fig:Res_cond_data}
The top (experimental) and bottom (theoretical) rows show absorption images of the MOT for $\tilde{\Omega}/2\pi = \SI{2.1}{MHz}$, $\delta/2\pi =\SI{-400}{kHz}$ and a variety of $\Delta$. }
\end{figure*}

As discussed in the main text, the light induced forces in the MOT are weak compared to gravity. Therefore the atoms in the MOT fall under gravity until the Zeeman shift, induced by the MOT quadrupole field, matches the MOT beam detuning. Due to the nature of the quadrupole field, the resulting resonance condition forms an elliptical ‘shell’ around the quadrupole centre, as shown by the contours of constant energy in Fig. \ref{fig:Res_cond}(a). The application of the dressing beam induces a spatially dependent AC Stark shift of the MOT transition, due to its Gaussian intensity profile. This leads to a distortion of the elliptical energy contours. The distortion is strikingly dependent of the sign of $\Delta$. Fig. \ref{fig:Res_cond}(b) and (c) show the energy contours for $\Delta<0$ and $\Delta>0$. For the case of $\Delta<0$, the resonance condition becomes shallower and `double-welled', leading to a spreading of the MOT and a weakening of the confinement. This can be seen in the top row of Fig. \ref{fig:Res_cond_data} where the MOT appears to form in two separate locations. This is the dominant reason for the difficulty in compensating the dressed MOT for $\Delta<0$. Conversely, for $\Delta>0$, the resonance condition becomes more deeply furrowed, facilitating strong confinement and making it easier to maintain a compensated MOT. This can also be seen in the top row of Fig. \ref{fig:Res_cond_data} where the MOT appears to form in lower positions, creating a vertically elongated MOT. 

Below the experimental data in Fig. \ref{fig:Res_cond_data} is the results of a two-dimensional Monte Carlo simulation, based upon \cite{Hanley2017}. The model is semi-classical and based upon the optical Bloch equations. From the steady-state solution of the optical Bloch equations, the scattering dynamics of the atoms are calculated. The addition of the dressing beam is included by solving the three-level optical Bloch equations analytically, using Mathematica. This gives rise to the AC Stark shift of the MOT transition as well as facilitating the inclusion of an atomic loss rate, where the probability of loss is given by 
\begin{equation}
P = \Gamma_r \rho_{rr} \delta t~,
\end{equation}  
where $\Gamma_r$ is the Rydberg state decay rate, $\rho_{rr}$ is the Rydberg state population and $\delta t$ is the simulation time-step. This simulates the event that the direct excitation of a Rydberg atom leads to the loss of that atom from the MOT. We observe excellent agreement between the theoretical and experimental spatial distributions of the atoms in the MOT, allowing us to include density dependent Rydberg-dressed interactions into our model.

\subsubsection{Adding Rydberg-dressed interactions} 

\begin{figure}
\centering
\includegraphics[width=\columnwidth]{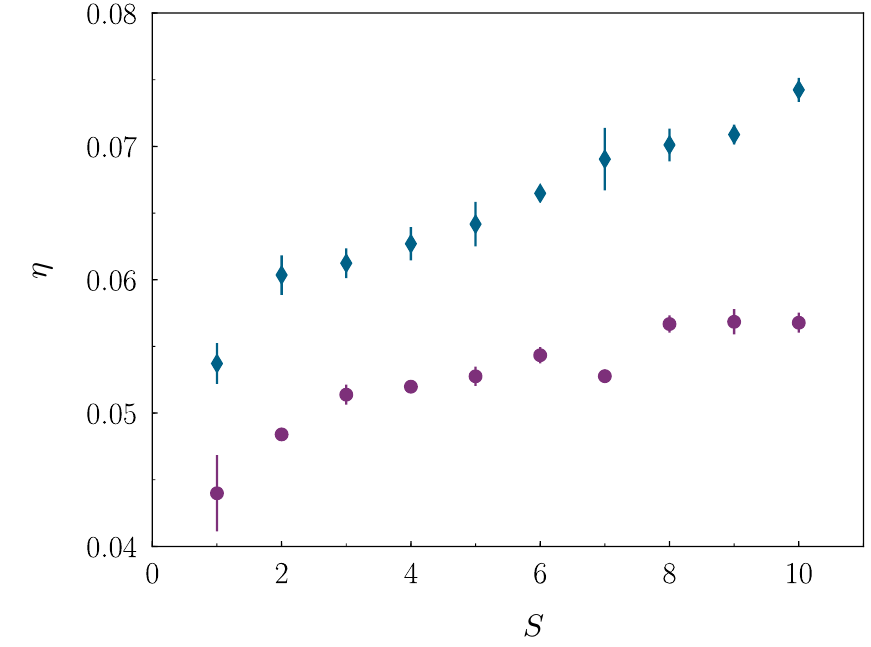}
\caption{\label{fig:rho_ee}
The figure shows the average excited fraction as a function of MOT beam power for beam powers in a ratio of 1:1:1 (blue diamonds) and 1:1:3 (red circles) where the ratio is between $\bm{\hat{x}}:\bm{\hat{y}}:\bm{\hat{z}}$ laser beam axes. The error bar shows the standard error on the mean.}
\end{figure}

Once the local density distribution of the atoms is known, it is possible to add density-dependent interactions to the model. However, as the Rydberg-dressed interactions only occur between atoms in the dressed state $\left|\tilde{e}\right\rangle$, we must first estimate $\eta$. Naively, one would expect that the steady-state, three-level Bloch solution would lead to an accurate value of $\rho_{ee}$. However, due to the motion of the atoms in the quadrupole field, this is not the case. We estimate $\eta$ by taking the average value of $\rho_{ee}$ calculated from the simulation. Fig. \ref{fig:rho_ee} shows $\eta$ as a function of $S$ for two experimental configurations. The first (blue diamonds) is where the power in all MOT beam directions is equal. The second (red circles) is where the power of the MOT beam in the vertical direction is three times that of the other MOT beams. We clearly observe a saturation effect at larger values of $S$ which arises from the frequency shift due to the atomic recoil following the absorption or emission of a photon.   

The Rydberg-dressed MOT operates in a complex regime where the interaction strength, dissipation and kinetic energy are all comparable in scale, limiting the usefulness of common approximations \cite{Macri2014,Genkin2014,Buchmann2017,Malossi2014,Marcuzzi2014,Weimer2010} for such many-body systems. To fully treat this interacting system, one must include correlations between atoms as the Rydberg-dressed interactions strongly correlates the scattering rate and hence the cooling of neighboring atoms. However, the coupling of correlations as well as atomic motion is a formidable theoretical challenge. For this reason, we consider a first-order mean-field approximation to the Rydberg-dressed potential. We approximate the Rydberg-dressed potential as  
\begin{align} 
V\left(N\left(\bm{r^{\prime}}\right),\bm{r}\right) &= \frac{N\left(\bm{r^{\prime}}\right)-1}{2}\eta V_0~&\rm{for}~&\left|\bm{r}-\bm{r^{\prime}}\right|<R_C~,\\ 
V\left(N\left(\bm{r^{\prime}}\right),\bm{r}\right) &= 0~&\rm{for}~&\left|\bm{r}-\bm{r^{\prime}}\right|>R_C~,
\end{align}
where $N\left(\bm{r^{\prime}}\right) = \pi R_{\rm{c}}^2 \rho_{\rm{2D}}\left(\bm{r^{\prime}}\right)$. This leads to an additional energy shift $V\left(N\left(\bm{r^{\prime}}\right)\right)/\hbar$ of $\left|\tilde{e}\right\rangle$ that depends on the local atomic density $\rho_{\rm{2D}}\left(\bm{r^{\prime}}\right)$. This additional energy shift is included in the model which in turn alters the scattering dynamics.  

\subsection{Radiation Pressure}
Here we detail radiation pressure effects in both the undressed and dressed MOT. The signatures of radiation trapping have been studied in a number of experiments \cite{Overstreet05,Townsend1995,Drewsen1994,Grego1996,Gabbanini1997,Vorozcovs05}, where the re-absorption of spontaneously emitted photons results in a repulsive force between the atoms which limits the maximum achievable density. 

For MOTs formed on inter-combination lines (as in our experiment) there is a suppression of this effect due to the reduced scattering rate compared to that of conventional MOTs, leading to larger achievable densities \cite{Katori1999}. However at high densities we observe an increase in both cloud temperature and width which can be attributed to radiation trapping. The radiation trapping force between two atoms separated by a distance $d$ is defined as \cite{Sesko1991,Foot1992}

\begin{equation}
F_{\rm{R}} = \frac{\sigma_{\rm{l}}\sigma_{\rm{r}}I}{4\pi cd^2}, 
\label{eqn:rad_press}
\end{equation}

where $\sigma_{\rm{l}}$ is the cross-section for absorption of a laser photon, $\sigma_{\rm{r}}$ a re-radiated photon and $I$ is the beam intensity. We define $\sigma_{\rm{l}}$ using the on resonance cross-section, $\sigma_{\rm{0}} = 3\lambda^{2}/2\pi$,
 
\begin{equation}
\sigma_{\rm{l}} = \frac{\sigma_{\rm{0}}}{1+S+4(\Delta/\Gamma)^{2}}. 
\end{equation}

\begin{figure}
\centering
\includegraphics[width=\columnwidth]{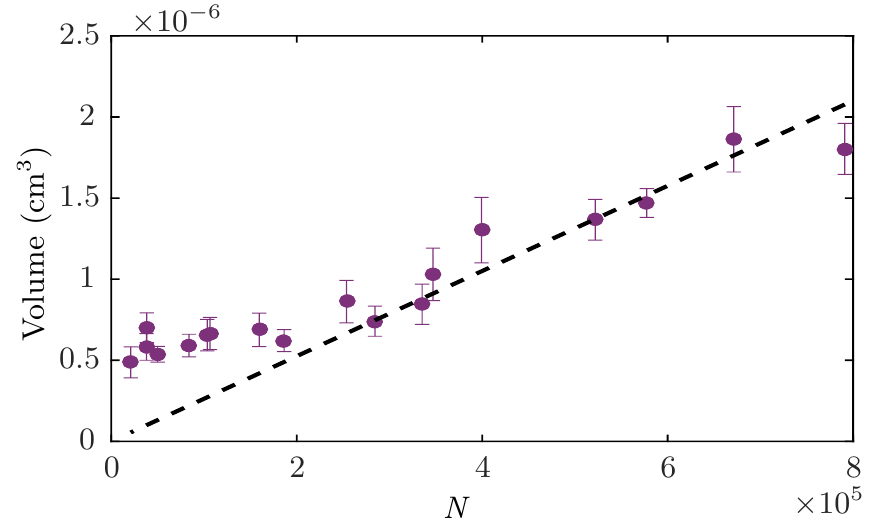}
\caption{\label{fig:radpress1} MOT volume versus the number of trapped atoms (black circles). Red dashed line is a fit through the origin for atoms numbers $>$ \num{2.5e5}. 
}
\end{figure}

From \cite{Sesko1991}, the ratio $\sigma_{\rm{r}}/\sigma_{\rm{l}}$ can be determined from $n_{\rm{ms}}$, the maximum attainable density in the multiple scattering regime. Due to the repulsive force between the atoms at higher trap densities, the MOT radius is determined by the number of atoms $N$, rather than the temperature, and scales as $N^{1/3}$. Fig.~\ref{fig:radpress1} shows the undressed MOT volume as a function of atom number. Here the increase in volume for atom numbers $> \SI{2.5e5}{}$ indicates the onset of radiation trapping. Fitting a straight line to the data in this regime predicts an $n_{\rm{ms}}=
\SI{1.1 \pm 0.6 e12}{cm^{-3}}$ and hence a cross-section ratio $\sigma_{\rm{r}}/\sigma_{\rm{l}} \sim 1$. An increase in temperature of \SI{0.3}{\upmu K} over a range $\rho$ of \SIrange[range-phrase = --,range-units = brackets]{1}{7e11}{cm^{-3}} provides further evidence of radiation trapping. This is in agreement with data shown in \cite{Katori1999}.

As shown in the main text, applying the dressing laser causes the cloud to expand and hence leads to a fast decay in density. The expansion results from a combination of radiation trapping along with a weakening in the trap confinement due to the Gaussian profile of the dressing beam, as described in the previous section. To first approximation, the instantaneous `switch-on' of the dressing beam means that the atoms do not initially strongly interact with the MOT beams due to the change in resonance condition (see Fig.~\ref{fig:Res_cond}). This implies that radiation pressure is the dominant force in the first $\sim\SI{2}{ms}$ and forces attributed to MOT beam scattering can be neglected. We simulate the initial expansion process of the MOT whilst dressing using a simple one-dimensional model which is based upon Eq. \ref{eqn:rad_press}. Initially we use a normal distribution of atoms with a $1/e^2$ radius of $\SI{60}{\upmu m}$, similar to the experimentally measured value. The total simulation is broken down into a series of time steps. At each time-step, we calculate the force and velocity of each atom due to the presence of all others. The atoms are then allowed to evolve following Newtonian dynamics, before the force is re-calculated. At each time step, the position of the atoms is recorded from which the width of the resulting distribution is measured. 

\begin{figure}
\centering
\includegraphics[width=\columnwidth]{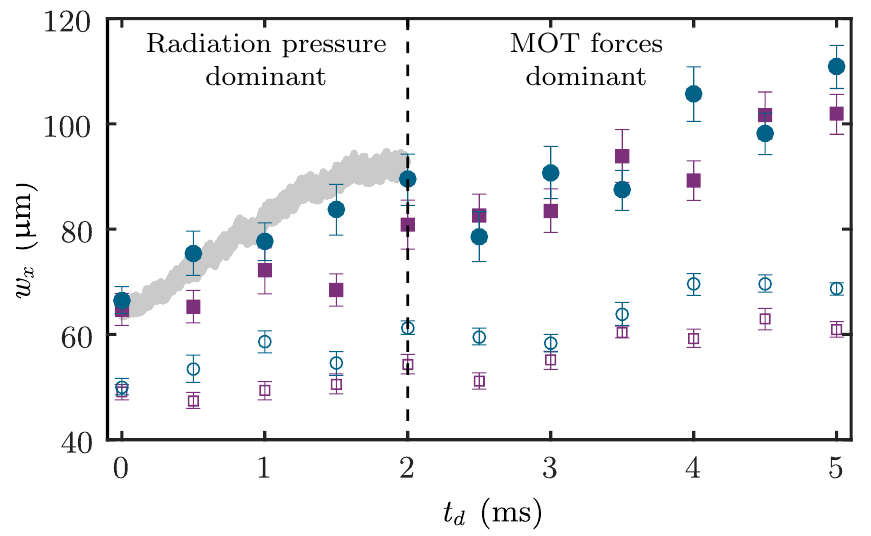}
\caption{\label{fig:rad_press2}
Dressed MOT width $w_x$ plotted as a function of dressing time $t_d$ for the $\rm{5s34d \ ^{3}D_{2}}$ (blue circles) and $\rm{5s37d \ ^{3}D_{2}}$ (purple squares) at high (filled markers) and low (hollow markers) initial densities. Also shown is the cloud width determined from the model (grey) as described in the text.
}
\end{figure}

Fig.\ref{fig:rad_press2} shows the model produces a similar expansion rate of the atoms in the first \SI{2}{ms}, compared to the experimentally measured widths $w_x$ of the high density MOT. For a low initial density the increase in width is much less significant, but at longer times as MOT forces become dominant we see similar gradients at both low and high densities. Furthermore, experiments taken at different states under the same conditions show that the expansion in cloud width is independent of the Rydberg state from $34<n<37$. These results provide evidence that the decay in density within the dressed MOT is due to a combination of radiation trapping and the change in resonance condition due to the non-uniform profile of the dressing beam. 

\bibliography{arXiv}

\begin{thebibliography}{67}
\bibitem [{For()}]{ForsterNote}%
  \BibitemOpen
  \href@noop {} {}\bibinfo {note} {The $n=37$ state is very close to a
  F\"{o}rster resonance leading to a large (and rather uncertain)
  $C_6$.}\BibitemShut {Stop}%
\end{thebibliography}

\end{document}